\documentclass[twoside,11pt]{article} 
\usepackage{amssymb}
\usepackage{fancyhdr}    % headings style
\newcommand{\Eins}{{\mathbf 1}} \newcommand{\RR}{{\mathbb R}}
\newcommand{\NN}{{\mathbb N}}
\newcommand{\HH}{{\cal H}} 
\newcommand{\be}{\begin{equation}} \newcommand{\ee}{\end{equation}}
\newcommand{\bea}{\begin{eqnarray}} \newcommand{\ea}{\end{eqnarray}}
\renewcommand{\theequation}{\arabic{section}.\arabic{equation}}
\renewcommand{\thesection}{\bf Lecture \arabic{section}:}

\newcommand{\e}{{\rm e}} 
\newcommand{\ads}{^{\rm AdS}} 
\newcommand{\adscft}{^{\hbox{\scriptsize AdS-CFT}}} 
\newcommand{\class}{_{\hbox{\scriptsize s-cl}}} 
\newcommand{\inv}{^{-1}}

\newcommand{\fmini}[1]{\vskip2mm
\framebox{\begin{minipage}{110mm}{\sl #1 \par}\end{minipage}}\vskip2mm}

\begin{document}
\baselineskip=11pt

\title{QFT Lectures on AdS-CFT \thanks{presented at the III.\
    Summer School in Modern Mathematical Physics, Zlatibor (Serbia and
    Montenegro), August 2004; supported by Deutsche Forschungsgemeinschaft}}
\author{\bf{K.-H. Rehren}\hspace{.25mm}\thanks{\,e-mail address:
rehren@theorie.physik.uni-goe.de}
\\ \normalsize{Inst.\ f\"ur Theoretische Physik, Univ.\ G\"ottingen}}

\date{}

\maketitle

\begin{abstract}
It is discussed to which extent the AdS-CFT correspondence is
compatible with fundamental requirements in quantum field theory. 
\end{abstract}

\vskip4mm
\noindent{\Large\bf Introduction} \vskip4mm

We reserve the term ``AdS-CFT correspondence'' for the field
theoretical model that was given by Witten \cite{W} and Polyakov et al.\
\cite{GKP} to capture some essential features of Maldacena's
Conjecture \cite{M}. It provides the generating functional for
conformally invariant Schwinger functions in $D$-dimensional Minkowski
space by using a classical action $I[\phi\ads]$ of a field on
$D+1$-dimensional Anti-deSitter space. In contrast to Maldacena's
Conjecture which involves string theory, gravity, and supersymmetric
large $N$ gauge theory, the AdS-CFT correspondence involves only
ordinary quantum field theory (QFT), and should be thoroughly
understandable in corresponding terms. 

In these lectures, we want to place AdS-CFT into the general context
of QFT. We are not so much interested in the many implications of
AdS-CFT \cite{AGMOO}, as rather in the question ``how AdS-CFT works''. 
We shall discuss in particular 
\begin{itemize}\itemsep0mm
\item why the AdS-CFT correspondence constitutes a challenge for
  orthodox QFT
\item how it can indeed be (at least formally) reconciled with the
  general requirements of QFT
\item how the corresponding (re)interpretation of the AdS-CFT
  correspondence matches with other, more conservative, connections
  between QFT on AdS and conformal QFT, which have been established
  rigorously.
\end{itemize}

The lectures are meant to be introductory. When we refer to rigorous
methods and results in QFT, our exposition never has the ambition of
being rigorous itself. We shall avoid the technical formulation of
almost all details, but nevertheless emphasize whenever such details are
crucial for some arguments, though often enough neglected.  

To prepare the ground, we shall in the first lecture remind the reader 
of some general facts about QFT (and its formal Euclidean functional
integral approach), with special emphasis on the passage between
real-time QFT and Euclidean QFT, the positivity properties which
are necessary for the probability interpretation of quantum theory,
and some aspects of large $N$ QFT.

Only in the second lecture, we turn to AdS-CFT, pointing out its apparent
conflict (at a formal level) with positivity. We resolve this conflict
by (equally formally) relating the conformal quantum field defined by
AdS-CFT with a limit of ``conventional'' quantum fields which does
fulfill positivity.   

The third lecture is again devoted to rigorous methods of QFT, which
become applicable to AdS-CFT by virtue of the result of the second
lecture, and which concern both the passage from AdS to CFT and the
converse passage. 

To keep the exposition simple, and in order to emphasize the extent to
which the AdS-CFT correspondence can be regarded as a
model-independent connection, we shall confine ourselves to bosonic
(mostly scalar) fields (with arbitrary polynomial couplings), and never
mention the vital characteristic problems pertinent to gauge (or
gravity) theories.

\section{QFT}
\setcounter{equation}{0}

A fully satisfactory (mathematically rigorous) QFT must fulfill a
number of requirements. These are, in brief: 
\begin{itemize}\itemsep0mm
\item Positive definiteness of the Hilbert space inner product.
\item Local commutativity of the fields% 
\footnote{We use the notation $\hat\phi$ in order to distinguish the real-time
{\em quantum} field (an operator[-valued distribution] on the Hilbert space)
from the Euclidean field $\phi_E$ (a random variable) and its
representation by a functional integral with integration variable
$\phi$, see below.} 
$\hat\phi$ at spacelike separation.
\item A unitary representation of the Poincar\'e group, implementing
  covariant transformations of the fields.
\item Positivity of the energy spectrum in one, and hence every inertial frame.
\item Existence (and uniqueness) of the ground state = vacuum $\Omega$.
\end{itemize}
Clearly, for one reason or another, one may be forced to relax one
or the other of these requirements, but there should be good physical
motivation to do so, and sufficient mathematical structure to ensure a safe
physical interpretation of the theory. E.g., one might relax the
locality requirement at very short distances where it has not been
tested directly, as long as macrocausality is maintained; or one might admit
modifications of the relativistic energy-momentum relation at very high
energies. But it is known that there are very narrow limitations
on such scenarios (e.g., \cite{P}, see also \cite{S} for a critical
discussion of the axioms). Hilbert space positivity may be absent at
intermediate steps, notably in covariant approaches to gauge theory,
but it is indispensable if one wants to saveguard the probabilistic
interpretation of expectation values of {\em observables}.  

The above features are reflected in the properties of the vacuum
expectation values of field products
\be W(x_1,\ldots,x_n)=(\Omega,\hat\phi(x_1)\ldots\hat\phi(x_n)\Omega),\ee
considered as ``functions'' (in fact, distributions) of the field
coordinates $x_i$, known as the Wightman distributions.
 
Local commutativity and covariance appear as obvious symmetry
properties under permutations (provided $x_i$ and $x_{i+1}$ are at
spacelike distance) and Poincar\'e transformations, respectively. The
uniqueness of the vacuum is a cluster property (= decay behaviour at
large spacelike separations). Further consequences for the Wightman
distributions will be described in the sequel.

\subsection{The Wick rotation}
The properties of Wightman functions allow for the passage to Euclidean
``correlation functions'', known as the ``Wick rotation''. Because
this passage and the existence of its inverse justify the most popular
Euclidean approaches to QFT, let us study in more detail what enters
into it. 

The first step is to observe that by the spectrum condition, the
Wightman distributions can be analytically continued to complex points
$z_i=x_i+iy_i$ for which $y_i-y_{i+1}$ are future timelike (the 
``forward tube''), by replacing the factors $\e^{-ik_i\cdot x_i}$ in
the Fourier representation by $\e^{-ik_i\cdot z_i}$. The reason is that
the momenta $k_i+\ldots+k_{n-1}+k_n$ (being eigenvalues of the 
momentum operator) can only take values in the future light-cone, so
that $\prod_i\e^{-ik_i\cdot z_i} = \e^{ik_n\cdot(z_{n-1}-z_n)} \cdot 
\e^{i(k_{n-1}+k_n)\cdot (z_{n-2}-z_{n-1})} \cdot 
\e^{i(k_{n-2}+k_{n-1}+k_n)\cdot(z_{n-3}-z_{n-2})} \cdot \ldots$ decay
rapidly if $z_{i}-z_{i-1}$ have future timelike imaginary parts,
and would rapidly diverge otherwise (i.e., outside the forward tube)
for some of the contributing momenta. The analytically continued
distributions are in fact analytic {\em functions} in the forward
tube. The Wightman {\em distributions} are thus boundary values (as
${\rm Im}\;(z_i-z_{i+1})\searrow 0$ from the future timelike
directions) of analytic Wightman {\em functions}. 

Together with covariance which implies invariance under the 
{\em complex} Lorentz group, the analytic Wightman functions can be
extended to a much larger complex region, the ``extended domain''. 
Unlike the forward tube, the extended domain contains real points which
are spacelike to each other, hence by locality, the Wightman functions
are symmetric functions in their complex arguments. This in turn
allows to extend the domain of analyticity once more, and one obtains
analytic functions defined in the Bargmann-Hall-Wightman domain. This 
huge domain contains the ``Euclidean points'' $z_i=(i\tau_i,\vec x_i)$ with 
real $\tau_i$, $\vec x_i$. Considered as functions of 
$\xi_i:=(\vec x_i,\tau_i)$, the Wick rotated functions are called the
``Schwinger functions'' $S_n(\xi_1,\dots\xi_n)$, which are symmetric,
analytic at $\xi_i\neq\xi_j$, and invariant under the Euclidean group. 

It is convenient to ``collect'' all Schwinger functions in a
generating functional  
\be S[j] := \sum\frac{1}{n!}\int\left(\prod d\xi_i\;j(\xi_i)\right) 
S_n(\xi_1,\dots\xi_n) \equiv \Big\langle \e^{\int
  d\xi\;\phi_E(\xi)j(\xi)}\Big\rangle.\ee  
Knowledge of $S[j]$ is equivalent to the knowledge of the Schwinger
functions, because the latter are obtained by variational derivatives,  
\be S_n(\xi_1,\dots\xi_n)
= \prod_i\frac{\delta}{\delta j(\xi_i)} S[j]\vert_{j=0}.\ee
The generating functional for the ``truncated (connected) Schwinger
functions'' $S_n^T(\xi_1,\dots\xi_n)$ (products of lower correlations
subtracted) is $S^T[j]=\log S[j]$. 

It should be emphasized that Fourier transformation, Lorentz
invariance, and energy positivity enter the Wick rotation in a crucial way, 
so that in general curved spacetime, where none of these features is
warranted, anything like the Wick rotation may by no means be expected
to exist. Hence, we have
\fmini{{\bf Lesson 1.} Euclidean QFT is a meaningful framework,
  related to some real-time QFT, only provided there is sufficient
  spacetime symmetry to establish the existence of a Wick rotation.}

AdS is a spacetime where the Wick rotation {\em can} be established.
Although a more global treatment is possible, pertaining also to QFT on a
covering of AdS \cite{BEM}, we present the core of the argument in a
special chart (the Poincar\'e coordinates), in which AdS appears as a
warped product of Minkowski spacetime $\RR^{1,D-1}$ with $\RR_+$. 

Namely, AdS is the hyperbolic surface in
$\RR^{2,D}$ given by $X\cdot X=1$ in the metric of $\RR^{2,D}$.
In Poincar\'e coordinates, 
\be X=\left(\frac z2 +\frac{1- x_\mu x^\mu}{2z},\frac{x^\mu}z, 
-\frac z2+\frac{1+ x_\mu x^\mu}{2z}\right) \qquad (z>0).\ee  
In these coordinates, the metric is
\be ds^2 = z^{-2}(\eta_{\mu\nu} dx^\mu dx^\nu - dz^2),\ee
hence for each fixed value of $z$, it is a multiple of the Minkowski
metric.

The group of isometries of AdS is $SO(2,D)$, which is also the
conformal group of Minkowski spacetime $\RR^{1,D-1}$. It contains a
subgroup $SO(1,D-1) \ltimes \RR^{1,D-1}$ preserving $z$ and
transforming the coordinates $x^\mu$ like the Poincar\'e group. The
rest of the group are transformations which act non-linearly on the
coordinates $z$ and $x$ in such a way that the boundary $z=0$ is
preserved, and its points $(z=0,x)$ transform like scale and special
conformal transformations of $x$. 

The natural spectrum condition on AdS requires positivity of the
generator of the timelike ``rotations'' in the two positive signature
directions of the embedding space. This generator turns out to be the
``conformal Hamiltonian'' $\frac12 (P^0+K^0)$, which is positive in a
unitary representation if and only if $P^0$ is positive. Hence, the
AdS spectrum condition is equivalent to the Poincar\'e spectrum
condition, and the Wightman distributions have analytic continuations
in the Poincar\'e forward tube. Furthermore, the complex AdS group
contains the complex Poincar\'e group, and local commutativity at
spacelike AdS distance (which is equivalent to $(x-x')_\mu(x-x')^\mu
-(z-z')^2<0$) entails local commutativity at spacelike Minkowski
distance $(x-x')_\mu(x-x')^\mu<0$. Therefore, by repeating the same
reasoning as in Sect.\ 1.1 for the variables $x^\mu$ only, the
Bargmann-Hall-Wightman domain of analyticity of the AdS Wightman functions
contains the points $(z_i,i\tau_i,\vec x_i)$ with real
$z_i,\tau_i,\vec x_i$. Writing $\xi=(\vec x,\tau)$ as before, these
``Euclidean points'' coincide with the points of Euclidean AdS  
\be \Xi=\left(-\frac z2+\frac{1-\vert\xi\vert^2}{2z},\frac{\xi^\mu}z,
  \frac z2+\frac{1+\vert\xi\vert^2}{2z}\right)\qquad (z>0),\ee
which satisfy $\Xi\cdot\Xi=1$ in the metric of $\RR^{1,D+1}$.

\subsection{Reconstruction and positivity}
By famous reconstruction theorems \cite{SW,OS}, the Wightman
distributions or the Schwinger functions completely determine the
quantum field, including its Hilbert space. For the reconstruction of
the Hilbert space, one {\em defines} the scalar product between
improper state vectors $\hat\phi(x_1)\ldots \hat\phi(x_n)\Omega$ to be
given by the Wightman distributions. This scalar product must be
positive: Let $P=P[\hat\phi]$ denote any polynomial in smeared
fields. Then one has  
\be (\Omega,P^*P\Omega)= \vert\vert P\Omega\vert\vert^2\geq 0.\ee 
Inserting the smeared fields for $P$, $(\Omega,P^*P\Omega)$ is a
linear combination of smeared Wightman distributions. Thus, every
linear combination of smeared Wightman distributions which can
possibly arise in this way must be non-negative. (It could be zero
because, e.g., $P$ contains a commutator at spacelike distance such
that $P=0$, or the Fourier transforms of the smearing functions avoid
the spectrum of the four momenta such that $P\Omega=0$.)   

This property translates, via the Wick rotation, into a property
called ``reflection positivity'' of the Schwinger functions:
Let $P=P[\phi_E]$ denote a polynomial in Euclidean fields smeared in a
halfspace $\tau_i>0$, and $\theta(P)$ the same polynomial smeared with
the same functions reflected by $\tau_i\mapsto-\tau_i$. Then 
\be \Big\langle \theta(P)^*P\Big\rangle \geq 0.\ee
This expression is a linear combination of smeared Schwinger
functions. Reflection positivity means that every linear combination
which can possibly arise in this way must be non-negative. 

As an example for the restrictivity of reflection positivity, we
consider the 2-point function of a Euclidean conformal scalar field of
scaling dimension $\Delta$,
$S_2(\xi_1,\xi_2)=\vert\xi_1-\xi_2\vert^{-2\Delta}$. Ignoring 
smearing, we choose $P[\phi_E]=\phi_E(\frac\tau 2,0)-\phi_E(\frac\tau
2,x)$ and obtain  
\be \Big\langle \theta(P)^*P\Big\rangle
=2\left[\tau^{-2\Delta}-(\tau^2+x^2)^{-\Delta}\right].\ee 
Obviously, this is positive iff $\Delta>0$. This is the unitarity
bound for conformal fields in 2 dimensions. (More complicated
configurations of Euclidean points in $D>2$ dimensions give rise to
the stronger bound $\Delta \geq \frac {D-2}2$.)

The positivity requirements (1.7) resp.\ (1.8) are crucial for the
reconstructions of the real-time quantum field, which start with the
construction of the Hilbert space by defining scalar products on
suitable function spaces in terms of Wightman or Schwinger functions
of the form (1.7) resp.\ (1.8). 

As conditions on the Wightman or Schwinger functions, the
positivity requirements are highly nontrivial. It is rather easy 
to construct Wightman functions which satisfy all the requirements
except positivity, and it is even more easy to guess funny Schwinger
functions which satisfy all the requirements except reflection
positivity. In fact, the remaining properties are only symmetry,
Euclidean invariance, and some regularity and growth properties, which
one can have almost ``for free''. 

But without the positivity, these functions are rather
worthless. From non-positive Wightman functions one would reconstruct
fields without a probability interpretation, and reconstruction from
non-positive Schwinger functions would not even yield locality and
positive energy, due to the subtle way the properties intervene in 
the Wick rotation. In particular, the inverse Wick rotation uses
methods from operator algebras which must not be relied on in
``Hilbert spaces'' with indefinite metric. 

\fmini{{\bf Lesson 2.} Schwinger functions without reflection positivity 
have hardly any physical meaning.}

\subsection{Functional integrals}
The most popular way to obtain Schwinger functions which are 
{\em at least in a formal way} reflection-positive, is via functional 
integrals \cite{GJ}: the generating functional is 
\be S[j]:=Z^{-1}\int D\phi\;\e^{-I[\phi]}
\cdot\e^{\int d\xi \,\phi(\xi)j(\xi)},\ee 
where $I[\phi]$ is a Euclidean action of the form
$\frac12(\phi,A\phi)+\int d\xi V(\phi(\xi))$ with a quadratic form $A$
(e.g., the Klein-Gordon operator) which determines a free propagator,
and an interaction potential $V(\phi)$. The normalization factor is
$Z=\int D\phi\;\e^{-I[\phi]}$. 

Varying with respect to the sources $j(\xi)$, the Schwinger functions are  
\be S_n(\xi_1,\dots\xi_n)
:=Z^{-1}\int D\phi\;\phi(\xi_1)\dots\phi(\xi_n)\;\e^{-I[\phi]},\ee
and one may think of them as the moments 
\be S_n(\xi_1,\dots\xi_n)
=\Big\langle\phi_E(\xi_1)\dots\phi_E(\xi_n)\Big\rangle,\ee
of random variables $\phi_E(\xi)$, such that the functional integration
variables $\phi$ are the possible values of $\phi_E$ with the 
probability measure $D\mu[\phi]=Z\inv D\phi\,\e^{-I[\phi]}$. 

The difficult part in constructing a Euclidean QFT along these lines is, 
of course, to turn the formal expressions (1.10) or (1.11) into well-defined
quantities \cite{J,GJ}. This task can be attacked in several different ways 
(e.g., perturbative or lattice approximations, or phase space cutoffs
of the measure) which all involve the renormalization of formally
diverging quantities. In the perturbative approach, the problems are
at least threefold: when one separates the interaction part from the
quadratic part of the action and writes $D\mu[\phi] \propto D\mu_0[\phi] 
\e^{-\int d\xi V(\phi(\xi))}$ where $D\mu_0[\phi]$ is a Gaussian
measure, and expands the exponential into a power series, then first,
$V(\phi)$ is not integrable with respect to $D\mu_0$ because it is not
a polynomial in {\em smeared} fields (UV problem); second, the $\xi$
integrations over $V(\phi(\xi))$ will diverge (IR problem); third, the
series will fail to converge. We shall by no means enter the
problem(s) of renormalization in these lectures, but we emphasize  

\fmini{{\bf Lesson 3.} The challenge of constructive QFT via
  functional integrals is to define the measure, in such a way
  that its formal benefits are preserved.} 

Among the ``formal benefits'', there is reflection positivity
which, as we have seen, is necessary to entail locality, energy
positivity, and Hilbert space positivity for the reconstructed
real-time field. It is not to be confused with the probabilistic
positivity property of the measure, which usually gets lost upon
renormalization, so that renormalized Schwinger functions in fact fail
to be the moments of a measure; but this latter property is not
required by general principles.  

Let us display the formal argument why the prescription (1.11)
fulfills reflection positivity. It consists in splitting  
\be \e^{-\int d\xi\; V(\phi(\xi))}=
\e^{-\int_{\tau<0}d\xi\; V(\phi(\xi))}\cdot
\e^{-\int_{\tau>0}d\xi \;V(\phi(\xi))} \equiv \theta(F)^*F\ee
with $\xi=(\vec x,\tau)$ and $F=F[\phi]= \e^{-\int_{\tau>0}d\xi
  \,V(\phi(\xi))}$. Then 
\be \Big\langle \theta(P)^*P\Big\rangle = 
\Big\langle \theta(FP)^*FP\Big\rangle_0\ee
where $\langle\ldots\rangle_0$ is the expectation value defined with
the Gaussian measure $D\mu_0[\phi]$, which is assumed to fulfill
reflection positivity. Viewing $F$ as an exponential series of smeared
field products, $\langle\theta(FP)^*FP\rangle_0$ and hence 
$\langle\theta(P)^*P\rangle$ is positive. We see that it is important
that the potential is ``local'' in the sense that it depends only on
the field at a single point, in order to allow the split (1.13) into
positive and negative Euclidean ``time''. 

Note that in gauge theories already the Gaussian measure $D\mu_0$
will fail to be reflection positive, a fact which has to be cured by
Gupta-Bleuler of BRST methods. 

Even with the most optimistic attitude towards Lesson 3 (``nothing
goes wrong upon renormalization''), we shall retain from Lesson 2 as a
guiding principle:
\fmini{{\bf Lesson 4.} A functional integral should not be trusted as
  a useful device for QFT if it violates reflection positivity already
  at the formal level.} 

\subsection{Semiclassical limit and large $N$ limit} 

For later reference, we mention some facts concerning the effect of
manipulations of generating functionals (irrespective how they are
obtained) on reflection positivity of the Schwinger
functions. 

The product $S[j]=S^{(1)}[j]S^{(2)}[j]$ of two (or more) reflection-positive
generating functional is another reflection-positive generating
functional. In fact, because the truncated Schwinger functions are
just added, the reconstructed quantum field equals
$\phi^{(1)}\otimes\Eins + \Eins\otimes\phi^{(2)}$ defined on
$\HH=\HH^{(1)}\otimes\HH^{(2)}$, or obvious generalizations 
thereof for more than two factors. In particular, positivity is
preserved if $S[j]$ is raised to a power $\nu\in\NN$. 

The same is not true for a power $1/\nu$ with $\nu\in\NN$: a crude way
to see this is to note that reflection positivity typically includes as
necessary conditions inequalities among truncated Schwinger $n$-point
functions $S^T_n$ of the general structure $S^T_4\leq S^T_2 S^T_2$,
while raising $S[j]$ to a power $p$ amounts to replace $S^T$ by 
$p \cdot S^T$.    

This remark has a (trivial) consequence concerning the semiclassical
limit: let us reintroduce the unit of action $\hbar$ and rewrite 
\be S[j]= Z^{-1}\int D\phi\;\e^{-\frac 1\hbar I[j;\phi]}\ee
where $I[j;\phi]=I[\phi]-\int \phi j$ is the action in the presence
of a source $j$. Appealing to the idea that when $\hbar$ is very
small, the functional integral is sharply peaked around the classical
minimum $\phi\class =\phi\class[j]$ of this action, let us replace
$\hbar$ by $\hbar/\nu$ and consider the limit $\nu\to\infty$. Then we
may expect (up to irrelevant constants) 
\be S\class[j] := \e^{-\frac 1\hbar I[j;\phi\class[j]]} = 
\lim_{\nu\to\infty} 
\left[\int D\phi\;\e^{-\frac \nu\hbar I[j;\phi]}\right]^{1/\nu}.
\ee
This generating functional treated perturbatively, gives the tree
level (semiclassical) approximation to the original one, all loop
diagrams being suppressed by additional powers of $\hbar/\nu$. 

The functional integral in square brackets is ``as usual'' with 
$\hbar/\nu$ in place of $\hbar$, hence we may assume that it satisfies
reflection positivity. But we have no reason to expect $S\class[j]$ to be
reflection-positive, because of the presence of the power
$1/\nu$. Thus $S\class[j]$ does not generate reflection-positive Schwinger
functions, and hence no acceptable quantum field. This is, clearly, no
surprise, because a classical field theory is not a quantum field
theory. 

A variant of this argument is less trivial, concerning the large
$N$ limit. If one raises $S[j]$ to some power $N$, the truncated
Schwinger functions are multiplied by the factor $N$, and diverge as
$N\to\infty$. Rescaling the field by $N^{-\frac12}$ stabilizes the
2-point function (assuming the 1-point function
$\langle\phi_E\rangle$ to vanish), but suppresses all higher truncated
$n$-point functions, so that the limit $N\to\infty$ becomes Gaussian,
i.e., one ends up with a free field. To evade this conclusion (the
``central limit theorem''), one has to ``strengthen'' the interaction at
the same time to counteract the suppression of higher truncated
correlations. Let us consider $S[j]$ of the functional integral form. 
Raising $S$ to the power $N$, amounts to integrate over $N$
independent copies of the field ($D^N\underline\phi=D\phi_1\ldots
D\phi_N$) with interaction $V(\underline\phi)=\sum_iV(\phi_i)$ and
coupling to the source $j\cdot \sum\phi_i$. One way to strengthen the
interaction is to replace, e.g., $V(\phi)=\lambda\sum_i\phi_i^4$ by
$V(\phi)=\lambda(\sum_i\phi_i^2)^2$ giving rise to much more
interaction vertices, coupling the $N$ previously decoupled copies of
the field among each other. At the same time, the action acquires an
$O(N)$ symmetry, so one might wish to couple the sources also only to $O(N)$
invariant fields, and replace the source term by
$j\cdot\sum\phi_i^2$, hence
\be I_N[j,\underline\phi] = \frac12 (\underline \phi,A\underline\phi) +
\int\lambda (\underline\phi^2)^2 + \int j\cdot\underline\phi^2.\ee
We call the resulting functional integral $S_N[j]$. 

All these manipulations maintain the formal reflection positivity of
$S_N[j]$ at any finite value of $N$. An inspection of the Feynman
rules for the perturbative treatment shows that now all truncated
$n$-point functions still carry an explicit factor of $N$, and
otherwise have a power series expansion in $N$ and $\lambda$ where
each term has less powers of $N$ than of $\lambda$. Introducing the 't
Hooft coupling $\theta=N\lambda$, this yields an expansion in $\theta$
and $1/N$. Fixing $\theta$ and letting $N\to\infty$, suppresses the
$1/N$ terms, so that the asymptotic behaviour at large $N$ is 
\be S_N[j]\sim \e^{N [S_\infty^T(\theta) + O(1/N)]}.\ee
To obtain a finite non-Gaussian limit, one has to take
\be S_\infty[j]:=\lim_{N\to\infty} S_N[j]^{1/N} = \e^{S_\infty^T(\theta)}.\ee
But this reintroduces the fatal power $1/N$ which destroys
reflection positivity. According to Lesson 4, this means
\fmini{{\bf Lesson 5.} The large $N$ limit of a QFT is {\em not}
itself a QFT.}
It is rather some classical field theory, for the same
reason as before: namely the explicit factor $N$ combines with the
tacit inverse unit of action $1/\hbar$ in the exponent of (1.18) to
the inverse of an ``effective'' unit of action $\hbar/N\to 0$. What
large $N$ QFT has to say about QFT, is the (divergent) asymptotic
behaviour of correlations as $N$ gets large.  

\section{AdS-CFT}
\setcounter{equation}{0}
\subsection{A positivity puzzle}
The AdS-CFT correspondence, which provides the generating functional for 
conformally invariant Schwinger functions from a classical action $I$ on 
AdS, was given by Witten \cite{W} and Polyakov {\em et al.}\ \cite{GKP} 
as a ``model'' for Maldacena's Conjecture. We shall discuss this
formula in the light of the previous discussions about QFT, in which
it appears indeed rather puzzling.  

First, the formula is essentially classical, because it is supposed to
capture only the infinite $N$ limit of the Maldacena conjecture. 
Its general structure is
\be S\adscft\class[j] := \e^{-I[\phi\ads[j]]}\ee
where $I[\phi\ads]$ is an AdS-invariant action of a field
on AdS, and $\phi\ads[j]$ is the (classical) minimum of the action $I$
under the restriction that $\phi\ads$ has prescribed boundary values
$j$. More precisely, introducing the convenient Poincar\'e coordinates
$(z>0,\xi\in\RR^D)$ of Euclidean AdS such that the boundary
$z=0$ is identified with $D$-dimensional Euclidean space, it is required
that the limit
\be (\partial\phi\ads)(\xi):=\lim_{z\to 0} z^{-\Delta}\phi\ads(z,\xi)\ee 
exists, and coincides with a prescribed function $j(\xi)$.

It follows from the AdS-invariance of the action $I[\phi\ads]$ (and
the assumed  
AdS-invariance of the functional measure) that the variational derivatives
of $S\adscft\class[j]$ with respect to the source $j$ are conformally 
covariant functions, more precisely, they transform like the correlation 
functions of a Euclidean conformal field of scaling dimension (``weight'') 
$\Delta$. Thus, symmetry and covariance are automatic. But how about 
reflection positivity?

To shed light on this aspect \cite{DR}, we appeal once more to the
idea that a functional integral is sharply peaked around the minimum
of the action, when the unit of action becomes small, and rewrite
$S[j]$ as 
\be S\adscft\class[j] = \lim_{\nu\to\infty} \Big[\int D\phi\ads
\e^{-\nu I[\phi\ads]}\cdot
\delta\left[\partial\phi\ads-j\right]\Big]^{1/\nu}\ee 
where a formal functional $\delta$-function restricts the integration to 
those field configurations whose boundary limit (2.2) exists and coincides
with the given function $j(\xi)$. We see that $\nu$ takes the role
of the inverse unit of action $1/\hbar$ in (2.3), so that
$\nu\to\infty$ signals the classical nature of this limit, hence of
the original formula.   

Now, there are two obvious puzzles concerning formal reflection positivity 
of this generating functional. The first is the same which was 
discussed in Sect.\ 1.4, namely the presence of the inverse power
$1/\nu$, which arises due to the classical nature of the formula.  
Even if the functional integral in square brackets were positive, this
power most likely would spoil positivity. (In fact, the correlation
functions obtained from $S\adscft\class$ can be seen explicitly to
have logarithmic rather than power-like short-distance singularities,
and hence manifestly violate positivity \cite{OK}.)  

The obvious cure (as it is of course also suggested in the original
papers \cite{W,GKP}) is to interpret the AdS-CFT formula (2.1) only as a
semiclassical approximation to the ``true'' (quantum) formula, and
consider instead the quantum version   
\be \left\langle \e^{\int d\xi\;\phi\adscft_E(\xi)j(\xi)}\right\rangle
\equiv S\adscft[j] := \int D\phi\ads\; \e^{-I[\phi\ads]}\cdot
\delta\left[\partial\phi\ads-j\right] \ee 
as the generating functional of conformally invariant Schwinger
functions of a Euclidean QFT on $\RR^D$. 

But the second puzzle remains: for this expression, the formal argument 
for reflection positivity of functional integrals, presented in Sect.\
1.3, fails: that argument treats the exponential of the interaction
part of the action as a field insertion in the functional integrand,
and it was crucial that field insertions $\phi$ in the functional 
integral amount to the same insertions of the random variable $\phi_E$
in the expectation value $\langle\ldots\rangle$, achieved by
variational derivatives of the generating functional $S$ with respect
to the source $j$. But this property (1.11) is not true for the AdS-CFT 
functional integral (2.4) where the coupling to the source is via
a $\delta$-functional rather than an exponential!

So why should one expect that the quantum AdS-CFT generating functional
satisfies reflection positivity, so as to be acceptable for a conformal QFT
on the boundary? Surprisingly enough, explicit studies of AdS-CFT
Schwinger functions, computing the operator product expansion
coefficients of the 4-point function at tree level \cite{OK}, show no
signs of manifest positivity violation which could not be restored in
the full quantum theory (i.e., regarding the logarithmic behaviour as
first order terms of the expansion of anomalous dimensions). Why is
this so?  

An answer is given \cite{DR} by a closer inspection of the Feynman
rules which go with the functional $\delta$ function in the
perturbative treatment of the functional integral. For simplicity, we
consider a single scalar field with quadratic Klein-Gordon action
$\frac12 \int\phi\ads(-\square +M^2)\phi\ads$ and a polynomial self-interaction. 
As usual, the Feynman diagrams for truncated $n$-point Schwinger
functions are connected diagrams with $n$ exterior lines attached to
the boundary points $\xi_i$, and with vertices according to the
polynomial interaction and internal lines connecting the vertices. 
Each vertex involves an integration over AdS. (For our considerations
it is more convenient to work in configuration space rather than in
momentum space.) However, the implementation of the functional 
$\delta$-function, e.g., by the help of an auxiliary field:
$\delta(\partial\phi\ads-j)=\int Db\, \e^{i\int b(\xi)
((\partial\phi\ads)(\xi)-j(\xi))}$, modifies the propagators. One has 
the bulk-to-bulk propagator $\Gamma(z,\xi;z',\xi')$ connecting two
vertices, the bulk-to-boundary propagator $K(z,\xi;\xi')$ connecting a
boundary point with a vertex, and the boundary-to-boundary propagator
$\beta(\xi;\xi')$ which coincides with the tree level 2-point function.   

The determination of these propagators is straightforward for a scalar
field, although the underlying ``general principles'' are somewhat subtle,
and will be described in some more detail in the appendix. The result is
the following. 

$\Gamma$ equals the Green function $G_+$ of the Klein-Gordon operator
which behaves $\sim z^{\Delta_+}$ near the boundary, where 
\be \Delta_\pm = \frac D2 \pm\sqrt{\frac{D^2}4+M^2}.\ee
It is a hypergeometric function of the Euclidean AdS distance.
$K$ is a multiple of the boundary limit $\lim_{z'\to 0}z'^{-\Delta_+} 
(\,\cdot\,)$ in the variable $z'$ of $G_+(z,\xi;z',\xi')$, and $\beta$
is a multiple of the double boundary limit in both variables $z$ and
$z'$ of $G_+$ \cite{BDHM}: 
\be \Gamma=G_+,\qquad K= c_1\cdot \lim_{z'\to 0}z'^{-\Delta_+}
G_+,\qquad \beta=c_2\cdot   
\lim_{z\to 0}z^{-\Delta_+} \lim_{z'\to 0}z'^{-\Delta_+} G_+\ee
with certain numerical constants $c_1$ and $c_2$. Specifically \cite{DR},  
\be c_1=2\Delta_+-D =\sqrt{D^2+4M^2},\ee
and, as will be crucial for the sequel, 
\be c_2=c_1^2.\ee

Now, let us consider the conventional (as in Sect.\ 1.3) functional
integral for a Euclidean field on AdS 
\be S\ads[J]=Z\inv\int D\phi\ads
\e^{-I[\phi\ads]}\e^{\int\sqrt g\;\phi\ads J\ads},\ee
choosing $G_+(z,\xi;z',\xi')$ as the propagator defining the Gaussian
functional measure. Its perturbative Schwinger functions are sums over
ordinary Feynman graphs with all lines given by $G_+$. Taking the
simultaneous boundary limits $\lim_{z_i\to 0}z_i^{-\Delta_+}(\,\cdot\,)$ 
of the Schwinger functions in all their arguments, one just has to
apply the boundary limit to the external argument of each external
line. This obviously yields bulk-to-bulk, bulk-to-boundary and
boundary-to-boundary propagators  
\be G_+,\qquad H_+= \lim_{z\to 0}z^{-\Delta_+} G_+,\qquad \alpha_+=
\lim_{z\to 0}z^{-\Delta_+} \lim_{z\to 0}z'^{-\Delta_+} G_+.\ee

Comparison of (2.6) and (2.10) implies for the resulting Schwinger functions
\be S\adscft_n(\xi_1,\ldots,\xi_n)=c_1^n\cdot
\left(\prod_i\lim_{z_i\to 0}z_i^{-\Delta_+}\right) 
S\ads_n(z_1,\xi_1,\ldots z_n,\xi_n)\ee
where it is crucial that $c_2=c_1^2$ because each external end of a
line must come with the same factor. 

In other words, we have shown that the Schwinger functions generated
by the functional integral (2.4) formally agree (graph by graph in
unrenormalized perturbation theory) with the boundary limits of those
generated by (2.9). The latter satisfy reflection positivity by the
formal argument of Sect.\ 1.3, generalized to AdS. Taking the joint
boundary limit preserves positivity, because this step essentially 
means a choice of special smearing functions (1.8), supported on the
boundary $z=0$ only. Thus, (2.4) indeed satisfies reflection 
positivity, in spite of its appearance. 

Because the Wick rotation affecting the Minkowski coordinates commutes
with the boundary limit in $z$, we conclude that the same relation
(2.11) also holds for the Wightman functions, and hence for the
reconstructed real-time quantum fields: 
\be \hat\phi\adscft(x) = c_1\cdot (\partial\hat\phi\ads)(x) \equiv c_1\cdot
\lim_{z\to 0}z^{-\Delta_+}\hat\phi\ads(z,x),\ee 
$x\in$ $D$-dimensional Minkowski spacetime. This relation describes
the restriction of an AdS covariant field to its timelike boundary
\cite{BBMS}, and generalizes the well-known fact that Poincar\'e
covariant quantum fields can be restricted to timelike hypersurfaces,
giving rise to quantum fields in lower dimensions, see Sect.\ 3.1. 
Moreover, because the AdS field (formally) satisfies reflection
positivity, so does its boundary restriction.   

We have established the identification (2.11), (2.12) for symmetric
tensor fields of arbitrary rank \cite{AG} (with arbitrary polynomial 
couplings), see the appendix. Although we have not considered 
antisymmetric tensors nor spinor fields, there is reason to believe
that this remarkable conclusion is true in complete generality. 

\fmini{{\bf Lesson 6.} Quantum fields defined by AdS-CFT are the boundary 
restrictions (limits) of AdS fields quantized conventionally on the
bulk (with the same classical action).}

We want to mention that in the semiclassical approximation (2.1), 
one has the freedom to partially integrate the classical
quadratic action and discard boundary contributions, which are of
course quadratic in $j$ and hence contribute only to the tree level
2-point function. This ambiguity has been settled previously
\cite{FMMR} by imposing Ward identities on the resulting correlation
functions. The resulting normalization $c_2$ of the tree level 2-point
function precisely matches the one obtained by the above functional method. 

Let us look at this from a different angle. Changing the tree level
2-point function amounts to multiplication of the generating
functional by a Gaussian. Thus, any different normalization would
add (as in Sect.\ 1.4) a Gaussian (free) field to the conformal
Minkowski field $\partial\hat\phi\ads$. Not surprisingly, the
sum would violate Ward identities which are satisfied by the field
without the extra Gaussian.

\section{Brane restrictions and AdS-CFT}
\setcounter{equation}{0}
We want to discuss the results obtained by formal reasoning in the
previous lecture, in the light of exact results on QFT.

\subsection{Brane restrictions}
Quantum fields may be restricted to timelike hypersurfaces \cite{B}. 
This is a non-trivial statement since they are distributions which
become operators only after smearing with smooth test functions, so it
is not obvious that one may fix one of the spacetime coordinates 
to some value. Indeed, $t=0$ fields in general do not exist in
interacting 4D theories. However, it is possible to fix one of the
spacelike coordinates thanks to the energy positivity, by doing so in
the analytically continued Wightman functions in the forward tube, 
which gives other analytic functions whose real-time limits 
${\rm Im}\;(z_i-z_{i+1})\searrow 0$ exist as distributions in a 
spacetime of one space dimension less.

The restricted field inherits locality (in the induced causal
structure of the hypersurface), Hilbert space positivity (because the
Hilbert space does not change in the process), and
covariance. However, only the subgroup which preserves the
hypersurface may be expected to act geometrically on the restricted
field. 

This result, originally derived for Minkowski spacetime \cite{B},
has been generalized to AdS in \cite{BBGMS}. Here, the warped product
structure implies that each restriction to a $z=$ const.\ hypersurface
(``brane'') gives a Poincar\'e covariant quantum field in Minkowski
spacetime. One thus obtains a family of such fields,
$\hat\phi_z(x):=\hat\phi\ads(z,x)$, defined on the same Hilbert space.  
Moreover, because spacelike separation in the Minkowski coordinates
alone implies spacelike separation in AdS, the fields of this family
are mutually local among each other. Even more, $\hat\phi_z(x)$ commute with
$\hat\phi_{z'}(x')$ also at timelike distance provided
$(x-x')_\mu(x-x')^\mu < (z-z')^2$. 

\subsection{AdS $\to$ CFT as QFT on the limiting brane}

Now assume in addition that the Wightman distributions $W\ads_n$ of a
(scalar) quantum field on AdS admit a finite limit 
\be \prod (\lim_{z_i\to 0} z_i^{-\Delta})
W\ads_n(z_1,x_1;\ldots;z_n,x_n) =: W_n(x_1,\ldots,x_n) \ee
for some value of $\Delta$. It was proven \cite{BBMS} that these
limits define a (scalar) Wightman field on Minkowski spacetime,
which may be written as 
\be \hat\phi(x)=(\partial\hat\phi\ads)(x) \equiv \lim_{z\to 0}z^{-\Delta}\hat\phi\ads(z,x). \ee
In addition to the usual structures, this field inherits conformal
covariance from the AdS covariance of $\hat\phi\ads$. It is an instructive
exercise to see how this emerges. 

Let $(z,x)\mapsto (z',x')$ be an AdS transformation (which acts
nonlinearly in these coordinates). This transformation takes (3.2)
into 
\be \lim_{z\to 0}z^{-\Delta}\hat\phi\ads(z',x')=
\lim_{z\to 0}(z'/z)^\Delta\cdot z'^{-\Delta}\hat\phi\ads(z',x').\ee
Now, because AdS transformations are
isometries, the measure $\sqrt{-g}\,dz\,d^Dx$ is invariant, where
$\sqrt{-g}=z^{-D-1}$. Hence
\be z^{-D-1} = z'^{-D-1} 
\det \left(\frac{\partial (z',x')}{\partial (z,x)}\right).\ee 
In the limit of $z\to 0$ (hence $z'\to 0$), $x'$ is a (nonlinear)
conformal transform of $x$. In the same limit, $\partial
z'/\partial x^\mu$ and $\partial x'^\mu/\partial z$ tend to $0$, and
$\partial z'/\partial z \approx z'/z$. Hence the Jacobian in (3.4) in that
limit becomes 
\be \det \left(\frac{\partial (z',x')}{\partial (z,x)}\right) \approx 
\frac{z'}z \cdot \det \left(\frac{\partial x'}{\partial x}\right).\ee
Hence, the factor $(z'/z)^\Delta$ in (3.3) produces the correct conformal
pre\-factors $\left(\det \left(\frac{\partial x'}{\partial x}\right)
\right)^{\frac\Delta D}$ required in the transformation law for
a scalar field of scaling dimension $\Delta$.  

None of the fields $\hat\phi_z$ ($z=$ const.\ $\neq 0$) is conformally
covariant because its family parameter $z$ sets a scale; hence the 
boundary limit may be re-interpreted as a scaling limit within a
family of non-scale-invariant quantum fields.

Comparing the rigorous formula (3.2) with the conclusion (2.12)
obtained by formal reasoning with unrenormalized perturbative
Schwinger functions, we conclude 
\fmini{{\bf Lesson 7.} The prescription for the AdS-CFT correspondence
coincides with a special instance of the general scheme of brane
restrictions, admitted in QFT.}

\subsection{AdS $\leftarrow$ CFT by holographic reconstruction}

In view of the preceding discussion, the inverse direction AdS
$\leftarrow$ CFT amounts to the reconstruction of an entire family of
Wightman fields $\hat\phi_z$ ($z\in\RR_+$) from a single member
$\hat\phi_{0} = \lim_{z\to 0}z^\Delta\hat\phi_z$ of that family, with
the additional requirement that two members of the family commute at
spacelike distance in AdS which involves the family parameters $z,
z'$. This is certainly a formidable challenge, and will not always be
possible. We first want to illustrate this in the case of a free
field, and then turn to a more abstract treatment of the problem in
the general case.    

Let us consider \cite{BBMS,GFF} a canonical Klein-Gordon field of mass
$M$ on AdS. The ``plane wave'' solutions of the Klein-Gordon equation
are the functions  
\be z^{D/2} J_\nu(z\sqrt {k^2}) \e^{\pm ik\cdot x}, \ee
where $\nu=\Delta-D/2=\sqrt{D^2/4+M^2}$, and the Minkowski momenta
range over the entire forward lightcone $V_+$. It follows that the
2-point function is 
\bea \langle\Omega,\hat\phi\ads(z,x)\hat\phi\ads(z',x')\Omega\rangle \sim
\qquad\qquad\qquad\qquad\qquad\qquad\qquad \nonumber \\ \sim (zz')^{D/2}
\int_{V_+} d^Dk J_\nu(z\sqrt {k^2})J_\nu(z\sqrt {k^2}) \e^{-ik(x-x')}
\sim \nonumber \\ \sim (zz')^{D/2} \int_{\RR_+} dm^2
J_\nu(zm)J_\nu(z'm) W_m(x-x')\ea 
(ignoring irrelevant constants throughout), where $W_m$ is the massive
2-point function in $D$-dimensional Minkowski spacetime. 

Restricting to any fixed value of $z$, we obtain the family of fields
$\hat\phi_z(x)$ which are all different ``superpositions'' of massive 
Minkowski fields with K\"allen-Lehmann weights $d\mu_z(m^2) = dm^2
J_\nu(zm)^2$. Such fields are known as ``generalized free fields''
\cite{J}. Using the asymptotic behaviour of the Bessel functions 
$J_\nu(u)\sim u^\nu$ at small $u$, the boundary field $\hat\phi_0$ turns
out to  have the K\"allen-Lehmann weight $d\mu_0(m^2) \sim m^{2\nu}dm^2$.  

In order to reconstruct $\hat\phi_z(x)$ from $\hat\phi_0(x)$, one has to 
``modulate'' its weight function, which can be achieved with the help
of a pseudo-differential operator:
\be \hat\phi_z(x) \sim z^\Delta \cdot j_\nu(-z^2\square)\hat\phi_0(x) \ee
where $j_\nu$ is the function $j_\nu(u^2)=u^{-\nu}J_\nu(u)$ on
$\RR_+$. Note that the operators $j_\nu(-z^2\square)$ are highly
non-local because $j_\nu(u)$ is not a polynomial, but they produce a
family of fields which all satisfy local commutativity with each other
at spacelike Minkowski distance \cite{GFF}.   

(In fact, the same is true for {\em any} sufficient regular function
$h(-\square)$, giving rise to an abundance of mutually local fields on
the same Hilbert space. The trick can also be generalized to Wick
products, by acting with operators of the form
$h(-\square_1,\dots,-\square_k)\vert_{x_1=\cdots=x_k}$. Moreover, 
although the generalized free field does not have a free Langrangean
and consequently no canonical stress-energy tensor, it does possess a
stress-energy tensor within this class of generalized Wick products,
whose $t=0$ integrals are the generators of conformal transformations.)

In order to reconstruct a local field $\hat\phi\ads(z,x)$ on AdS 
which fulfils local commutativity with respect to the causal structure
of AdS, Minkowski locality is, however, not sufficient. A rather
nontrivial integral identity for Bessel functions guarantees that
$\hat\phi_z(x)$ and $\hat\phi_{z'}(x')$ commute even at timelike distance
provided $(x-x')_\mu(x-x')^\mu < (z-z')^2$. Only this ensures that 
$\hat\phi\ads(z,x):=\hat\phi_z(x)$ is a local AdS field.

We have seen that the reconstruction of a local AdS field from its
boundary field is a rather nontrivial issue even in the case of a free
field, and exploits properties of free fields which are not known how
to generalize to interacting fields.  

In the general case, there is an alternative algebraic reconstruction
\cite{AH} of local AdS observables, which is however rather abstract
and does not ensure that these observables are smeared fields in the
Wightman sense. This approach makes use of 
the global action of the conformal group on the Dirac completion of
Minkowski spacetime, and of a corresponding global coordinatization
of AdS (i.e., unlike most of our previous considerations, it does not
work in a single Poincar\'e chart $(z,x)$). 

The global coordinates of AdS are
\be X= (\frac 1{\cos\rho}\vec e,\frac{\sin\rho}{\cos\rho}\vec E) \ee
where $\rho<\frac\pi 2$ and $\vec e$ and $\vec E$ are a 2-dimensional
and a $D$-dimensional unit vector, respectively. A parametrization of
the universal covering of AdS is obtained by writing $\vec
e=(\cos\tau,\sin\tau)$ and considering the timelike coordinate
$\tau\in\RR$. Thus, AdS appears as a cylinder $\RR\times
B^D$. While the metric diverges with an overall factor $\cos^{-2}\rho$ 
with $\rho\nearrow\frac\pi 2$ as the boundary is reached, lightlike
curves hit the boundary at a finite angle.

The boundary manifold has the structure of $\RR\times S^{D-1}$, which
is the universal covering of the conformal Dirac completion of
Minkowski spacetime. 

We consider causally complete boundary regions $K\subset \RR\times S^{D-1}$, 
and associate with them causally complete ``wedge'' regions
$W(K)\subset \RR\times B^D$, which are the causal completion of $K$ in
the causal structure of the bulk. It then follows that $W(K_1)$ and
$W(K_2)$ are causal complements in the bulk of each other, or AdS
transforms of each other, iff $K_1$ and $K_2$ are causal complements
in the boundary of each other, or conformal transforms of each other,
respectively.  

Now, we assume that a CFT on $\RR\times S^{D-1}$ is given. We want to
define an associated quantum field theory on AdS. Let $A(K)$ be the
algebras generated by CFT fields smeared in $K$. Then, by the
preceding remarks, the operators in $A(K)$ have the exact properties
as to be expected from AdS quantum observables localized in $W(K)$,
namely AdS local commutativity and covariance. AdS observables
in compact regions $O$ of AdS are localized in every wedge which
contains $O$, hence it is consistent to define \cite{AH}
\be A\ads(O):= \bigcap_{W(K)\supset O} A(K) \ee
as the algebra of AdS observables localized in the region $O$. Because
any two compact regions at spacelike AdS distance belong to some
complementary pair of wedges, this definition in particular guarantees
local commutativity. Note that this statement were not true, if only
wedges within a Poincar\'e chart $(z,x)$ were considered.

\fmini{{\bf Lesson 8.} Holographic AdS-CFT reconstruction is possible in
  general without causality paradoxes, but requires a global
  treatment.} 

The only problem with this definition is that the intersection of
algebras might be trivial (in which case the QFT on AdS has only
wedge-localized observables). But when the conformal QFT on the
boundary arises as the restriction of a bulk theory, then we know that
the intersection of algebras (3.10) contains the original bulk field
smeared in the region $O$.

\subsection{Conformal perturbation theory via AdS-CFT}
As we have seen, a Klein-Gordon field on AdS gives rise to a
generalized free conformal field. Perturbing the former by an
interaction, will perturb the latter. But perturbation theory of a
generalized free field is difficult to renormalize, because there is a
continuum of admissible counter terms associated with the continuous
K\"allen-Lehmann mass distribution of the generalized free field.

This suggests to perform the renormalization on the bulk, and then
take the boundary limit of the renormalized AdS field. Preserving AdS
symmetry, drastically reduces the free renormalization
parameters. 

This program is presently studied \cite{PTGFF}. Two observations are
emerging: first, to assume the existence of the boundary limit of the 
remormalized AdS field constitutes a nontrivial additional renormalization
condition; and second, the resulting renormalization scheme for the
boundary field differs from the one one would have adopted from a
purely boundary (Poincar\'e invariant) point of view. 

We do not enter into this in more detail \cite{PTGFF}. Let us just 
point out that this program can be successful only for very special
interactions of the conformal field, which ``come from AdS''. To
illustrate what this means, let us rewrite a typical interaction
Lagrangean on AdS as an interaction of the conformal boundary field,
using the results of Sect.\ 3.3: 
\bea \int \sqrt{-g}dz\,d^Dx\;\phi(z,x)^k = 
\int z^{-D-1}dz\,d^Dx\;\Big(z^\Delta j_\nu(-z^2\square)\phi_0(x)\Big)^k
=\nonumber \\ 
=\int d^Dx \Bigg(\int z^{k\Delta-D-1}dz\prod_{i=1}^k j_\nu(-z^2\square_i)\Bigg)
\prod_{i=1}^k \phi_0(x_i)\vert_{x_1=\cdots=x_k=x}.\quad  \ea
Reading the last expression as $\int d^Dx\;{\cal L}[\phi_0](x)$, one
encounters a conformal interaction potential ${\cal L}[\phi_0]$ which
involves another highly non-local pseudo-differential operator  
$\int z^{k\Delta-D-1}dz\prod_{i=1}^k
j_\nu(-z^2\square_i)(\,\cdot\,)\vert_{x_1=\cdots =x_k=x}$ acting on a
field product. It is crucial that this operator gives a local field
(i.e., when applied to the normal ordered product
${\colon}\hat\phi_0^k{\colon}$ of the quantum generalized free field,
the resulting field ${\cal L}[\hat\phi_0]$ satisfies local
commutativity with respect to $\hat\phi_0$ as well as with respect to
the family $\hat\phi_z$ and to itself), because otherwise the
interaction would spoil locality of the interacting field. In fact,
${\cal L}[\hat\phi_0]$ belongs to the class of generalized Wick
products mentioned in Sect.\ 3.3.

\appendix
\section{Appendix: AdS-CFT propagators}
\setcounter{equation}{0}
\renewcommand{\theequation}{\thesection.\arabic{equation}}
Because the chain of arguments leading to the Feynman rules for (2.4)
and to the validity of (2.8) (which together ultimately lead to (2.11)) 
is somewhat subtle \cite{DR}, we give here a more detailed outline. 
Moreover, we present the generalization to symmetric tensor fields
which was not published before \cite{AG}. 

The AdS-CFT propagators $\Gamma$, $K$, and $\beta$ in Sect.\ 2 are
determined as follows. First, we note that the Klein-Gordon equation
dictates the $z$-behaviour of its solutions near the boundary to be
proportional to $z^\Delta$ where $\Delta$ is related to the
Klein-Gordon mass by $\Delta(\Delta-D)=M^2$. There are thus two
possible values  
\be \Delta_\pm = \textstyle\frac 12 (D \pm\sqrt{D^2+4M^2}),\ee
and two Green functions $G_\pm(z,\xi;z',\xi')$ \cite{BF} which go like
$(zz')^{\Delta_\pm}$ as $z,z'\to0$. $G_\pm$ are hypergeometric
functions of the Euclidean AdS distance. Choosing either $G_+$ or
$G_-$ as a ``bare'' propagator, may be considered as the definition of
the Gaussian functional measure on which the perturbation series is
based. However, the diagrammatic bulk-to-bulk propagator $\Gamma_\pm$
differs from the bare propagator due to the presence of the functional
$\delta$ function. This can be seen, e.g., by implementing the
$\delta$-function by the help of an auxiliary field,
$\delta(\partial\phi\ads-j)=\int Db \e^{i\int b(\xi)
((\partial\phi\ads)(\xi)-j(\xi))}$, which introduces additional
quadratic terms. $\Gamma_\pm$ should still be a Green function, but
vanish faster than the bare propagator, which comes about as a
``Dirichlet condition'' due to the prescribed boundary values in the
functional integral. Because $\Delta_+>\Delta_-$, only $\Gamma_-$
exists (when the bare propagator is $G_-$), and coincides with
$G_+\sim z^{\Delta_+}$. For the other choice of $\Delta$, the
functional $\delta$-function cannot be defined. 

The bulk-to-boundary propagator is of group-theoretic origin \cite{D}. 
Name\-ly, the isometry group of $D+1$-dimensional Euclidean AdS and the 
conformal group of $\RR^D$ both coincide with $SO(D+1,1)$. The
solutions to the Klein-Gordon equation on AdS carry a representation
of the AdS group. Taking the boundary limit 
$(\partial_\pm\phi_\pm)(\xi):=\lim_{z\to 0}z^{-\Delta_\pm}\phi(z,\xi)$
of the solutions with either power law, one obtains functions on
$\RR^D$ which transform under $SO(D+1,1)$ like conformal fields of
dimension $\Delta$. The bulk-to-boundary propagator $K_\pm(z,\xi;\xi')$ is now an intertwiner between these representations, i.e.,  
\be \phi_\pm(z,\xi)= \int K_\pm(z,\xi;\xi')f(x') d^Dx\ee 
is a solution which transforms like a scalar field if $f$ transforms
like a conformal field of dimension $\Delta_\pm$. This property
determines $K_\pm$ to be proportional to 
\be K_\pm(z,\xi;\xi')\sim 
\left(\frac z{z^2+\vert\xi-\xi'\vert^2}\right)^{\Delta_\mp}. \ee
The absolute normalization of $K_\pm$ is given by the requirement that 
the boundary limit $(\partial_\pm\phi_\pm)(\xi)$ of (A.2) is
again $f(\xi)$; in other words, 
\be \lim_{z\to 0}z^{-\Delta_\pm} K_\pm(z,\xi;\xi')= 
\delta(\xi-\xi').\ee 
The boundary limit of the right hand side of (A.3) is a multiple of
$\delta(\xi-\xi')$ for the lower sign, just because
$\Delta_+>\Delta_-$ and $\Delta_++\Delta_-=D$, while it 
diverges for the other sign. Thus, only the bulk-to-boundary
propagator $K_-$ exists, while $K_+$ for the other choice of 
the Gaussian measure, like $\Gamma_+$, is ill defined. 

Finally, the tree level 2-point function $\beta_\pm(\xi,\xi')$
is found to be 
\be \beta_\pm=-\alpha_\pm\inv \ee 
where $\alpha_\pm = \partial_\pm\partial_\pm'G_\pm$ is the boundary
limit in both variables of the bare Green function $G_\pm$. The
inverse is understood as an integral kernel. A simple scaling argument
shows that these double boundary limits are proportional to 
$\vert\xi-\xi'\vert^{-2\Delta_\pm}$, and their inverses are  
\be \beta_\pm(\xi,\xi') \sim \vert\xi-\xi'\vert^{-2\Delta_\mp}. 
\ee

By inspection of these explicit functions, one finds \cite{BDHM}
\be \Gamma_-=G_+,\qquad K_-= c_1\cdot \partial'_+G_+,\qquad \beta_-=c_2\cdot   
\partial_+\partial'_+ G_+\ee
with numerical constants $c_1$ and $c_2$, to be determined below.

All the arguments given above for the scalar case generalize 
{\em mutatis mutandis} to the case of symmetric tensor fields of
arbitrary rank \cite{AG}. For group-theoretical reasons, one always has 
\be \Delta_++\Delta_-=D. \ee 
Namely, for each tensor rank $r$, the covariant Klein-Gordon equation
is in fact an eigenvalue equation \cite{F} for the quadratic Casimir
operator of the isometry group $SO(D+1,1)$ of AdS, $C=M^2+r(r+D-1)$,
while in the conformal interpretation of the same representation,
$C=\Delta(\Delta-D)+r(r+D-2)$. Equating the two eigenvalues
\be \Delta(\Delta-D)=M^2+r,\ee 
one obtains two solutions $\Delta_\pm$ related by (A.8). 

That $K_-$ and $H_+=\partial'_+G_+$ (the boundary limit of $G_+$)
are proportional to each other for any rank $r$, 
\be K_-=c_1\cdot H_+.\ee
follows because the inter\-twining
property of $K_-$ and the definition of $H_+$ as a limit of a Green
function lead to the same linear differential equations for both functions, 
with the same symmetry and boundary conditions. More precisely, 
both are ``bi-tensors'' (with AdS resp.\ Euclidean
$r$-fold multi-indices $\underline A$ and $\underline a$) 
subject to the homogeneous conditions 
\begin{itemize}\itemsep0mm
\item AdS- and conformal covariance (with weight $\Delta=\Delta_+=$ the
  larger solution of (A.9)) under simultaneous transformations of
  $(z,\xi)$ and $\xi'$, entailing homogeneity in all variables. 
\item symmetry and vanishing trace both as an AdS and a Euclidean
  tensor.
\item vanishing covariant divergence $D^{A_i} X_{\underline
    A;\underline a}=0$.
\item Klein-Gordon equation $(-D^CD_C +M^2) X_{\underline
    A;\underline a}=0$. 
\end{itemize} 
These conditions uniquely determine the structure
\be X_{\underline A;\underline a}(z,\xi;\xi')=v^{r-\Delta}
\left[\prod_{i=1}^r\left(D_{A_i}\partial'_{a_i} \log v\right)
\right]_{\rm symm} - \hbox{contractions}\cdot \delta_{a_ia_j} \ee
up to normalization, hence $(K_-)_{\underline A;\underline a}$ and
$(H_+)_{\underline A;\underline a}$ are both multiples of 
$X_{\underline A;\underline a}$. (For a sketch of the proof, see
below; cf.\ also \cite{D} for a group-theoretical derivation.) Here  
\be v=v(z,\xi;\xi') = \frac{z^2+\vert\xi-\xi'\vert^2}{2z} 
=\lim_{z'\to0} z'u(z,\xi;z',\xi')\ee
where $u=\frac{(z-z')^2+\vert\xi-\xi'\vert^2}{2zz'}=2\sinh^2s$ is
a function the geodesic distance $s$ on Euclidean AdS. The
contractions render $X_{\underline A;\underline a}$ traceless in
the boundary indices.

The intertwiner $K_-$ is normalized by the generalization of (A.4),
\be
\lim_{z\to0} \int d\xi' z^{r+\Delta-D}(K_-)_{\underline b;\underline
    a}(z,\xi;\xi')f_{\underline a}(\xi') = f_{\underline b}(\xi)
\ee
for any symmetric and traceless smearing function $f_{\underline a}(\xi)$.
On the other hand, $H_+$ being the boundary limit of a Green function
is normalized by  
$$ 
\int\! \sqrt g\, dz\, d\xi\; [(-D^CD_C+M^2) F^{\underline A}(z,\xi)]
(H_+)_{\underline A;\underline a}(z,\xi;\xi') = \lim_{z'\to
  0}z'^{r-\Delta}F_{\underline a}(z',\xi'), 
$$
for any symmetric, traceless and covariantly conserved smearing function
$F_{\underline A}(z,\xi)$. Choosing the boundary components of this
function of the form $F_{\underline a}(z,\xi)=z^{\Delta-r}
f_{\underline a}(\xi)$, and all $z$-components $=0$, this condition
reduces to 
\be 
\int z^{r+\Delta-D+1}dz\int d\xi\; [-\square f_{\underline b}(\xi)]
(H_+)_{\underline b;\underline a}(z,\xi;\xi') = f_{\underline a}(\xi')
\ee
for any symmetric and traceless smearing function $f_{\underline a}(\xi)$.

From this, we obtain the absolute normalizations of $K_-$ and
$H_+$ separately (see below), and then determine the relative
coefficient in (A.10). We find, universally for {\em every} tensor rank
$r=0,1,2,\ldots$  
\be c_1(\Delta)=2\Delta-D. \ee

For the matching condition (2.8), we have to compute also $c_2$. 
This can be done by a purely structural argument: Let $H_\pm$ and
$\alpha_\pm$ be the respective boundary limits of the Green functions
$G_\pm$ in one and in both variables. Then    
\be K_- = H_-\cdot \alpha_-\inv \ee
formally fulfills the required properties of the bulk-to-boundary
propagator including the normalization conditon (A.4) or its
generalization (A.13) that its boundary limit is the $\delta$-function. 
On the other hand, $K_-=c_1(\Delta_+)\cdot H_+$, hence 
\be c_1(\Delta_+)\cdot H_+\alpha_-=H_-.\ee
If we knew the analogous identity for the opposite signs, 
\be \tilde c_1\cdot H_-\alpha_+=H_+, \ee 
then we could conclude $c_1\tilde c_1\cdot H_-\alpha_+=K_-$ and, applying the 
boundary limit to both sides, $c_1\tilde c_1\cdot \alpha_-\alpha_+=\delta$, 
hence $\alpha_-\inv=c_1\tilde c_1\cdot \alpha_+$. Because 
$\beta_-=-\alpha_-\inv$ (see above), we would conclude $c_2=-c_1\tilde c_1$ 
in (A.7). 

The problem is, that the intergration in (A.18) is UV-divergent and
has to be regularized. Using the fact that $H_\pm$ and $\alpha_\pm$
are the values of analytic functions $H(\Delta)$ and $\alpha(\Delta)$
at the points $\Delta=\Delta_\pm$, and (A.17) is true in an open
region of the complex variable $\Delta$, we regularize (A.18) by
analytic continuation from (A.17). This implies, that (A.18) is valid
with $\tilde c_1$ the value of the analytic function $c_1(\Delta)$ at the
point $\Delta_-$, and hence
\be c_2=-c_1(\Delta_+)\cdot c_1(\Delta_-). \ee

The matching condition (2.8) is thus equivalent to the symmetry 
\be c_1(\Delta_+)+c_1(\Delta_-)=0,\ee
which is indeed satisfied by the function $c_1=c_1(\Delta)$ in (A.15).

Although we have not considered antisymmetric tensors nor spinor
fields, the universality of (A.15) makes one believe that the
remarkable conclusion (2.11), (2.12) is true in complete generality. 

Let us now turn to proving (A.11) and (A.15). The bi-tensor
$X_{\underline A;\underline a}$ satisfies the required covariance
properties because $v$ is given by the scaled limit (A.12) of the invariant
distance $u$. It is traceless as a Euclidean tensor by construction. 
Contracting with $g^{A_iA_j}$, the identities 
\be (D^A v)(D_A v) = v^2 \quad \hbox{and}\quad (D^A\partial'_a v)
(D_A\partial'_b v)=\delta_{ab}+ (\partial'_a v)(\partial'_b v)\ee
imply that $(D^A\partial'_{a} \log v) (D_{A}\partial'_{b} \log v)=
\delta_{ab}v^{-2}$, so the contributions from the displayed leading term 
in (A.11) cancel against the contractions because we know that the whole 
is traceless as a Euclidean tensor. Hence $X_{\underline A;\underline a}$ 
is automatically also traceless as an AdS tensor. Similarly, a covariant
divergence of the displayed term of (A.11) involves terms
$(D^Av)(D_{A}\partial'_{a}\log v)$ and $D^AD_{A}\partial'_{a}\log v$
which both vanish due to (A.21), as well as terms
$(D^{A}D_B\partial'_{b}\log v)(D_{A}\partial'_{a}\log v)
+(a\leftrightarrow b)= D_B(\delta_{ab}v^{-2})$ which again cancel 
against the contractions.

Computing the covariant Laplacian $D^CD_C$ acting on the displayed term
of (A.11), one gets contributions involving either $D^CD_C v^{r-\Delta}$ 
or $D^CD_C(D_{A}$ $\partial'_{a}\log v)$ or 
$(D^Cv^{r-\Delta})(D_CD_{A}\partial'_{a}\log v)$ or 
$(D^CD_{A}\partial'_{a}\log v)(D_CD_{B}\partial'_{b}\log
v)+(a\leftrightarrow b)$. Using (A.21) and the identity
\be D_AD_Bv=g_{AB}\cdot v,\ee
each of these contributions turns out to be a multiple of the displayed
term itself, the last one with an additional term involving
$\delta_{ab}$. The multiples sum up to $\Delta(\Delta-D)-r =
M^2$. Hence, the Klein-Gordon equation is fulfilled up to terms
involving $\delta_{ab}$, which we know to cancel among each other as
before. This proves the correctness of the structure (A.11).

Now, in order to determine the absolute normalizations from (A.13) and
(A.14), one only has to insert the structure 
$X_{\underline A;\underline a}$ and perform the integrals. The
contraction terms do not contribute. The crucial step is to rewrite
one factor 
$z\,v^{2-r-\Delta}\,(\partial_a\partial_b\log v)$ appearing in
each of these integrals, as   
$$\textstyle \frac{r+\Delta-2}{r+\Delta-1}v^{1-r-\Delta}\delta_{ab}+ \frac
1{r+\Delta-1} \partial_b(\xi_av^{1-r-\Delta})$$
and then perform a partial integration with the second term. 
The contributions from the partial integration vanish by symmetry and
tracelessness of the smearing functions, using the limit $z\to0$ in the
case of (A.13) and by the vanishing of the divergence in the case of
(A.14). Hence in both cases the rank $r$ integral is reduced to the
corresponding rank $r-1$ integral with an additional factor
$\frac{r+\Delta-2}{r+\Delta-1}\delta_{ab}$. Thus, the same factors 
enter the absolute normalizations upon passage from rank $r-1$ to $r$,
leaving the relative normalization independent of the rank. Thus (2.7)
computed once for the scalar field, gives (A.15) for any rank.


\begin{thebibliography}{B-B} %%%%%
\medskip\itemsep0.4mm
\begin{footnotesize} 

\bibitem{AGMOO}
O.~Aharony, S.~S.~Gubser, J.~M.~Maldacena, H.~Ooguri and Y.~Oz,
{\it Large $N$ field theories, string theory and gravity},
Phys.\ Rept.\ {\bf 323} (2000) 183--386.
%%CITATION = HEP-TH 9905111;%%


\bibitem{BDHM}
T.~Banks, M.~R.~Douglas, G.~T.~Horowitz and E.~J.~Martinec,
{\it AdS dynamics from conformal field theory},
arXiv:hep-th/9808016.
%%CITATION = HEP-TH 9808016;%%

\bibitem{BBGMS}
M.~Bertola, J.~Bros, V.~Gorini, U.~Moschella and R.~Schaeffer,
{\it Decomposing quantum fields on branes},
Nucl.\ Phys.\ B {\bf 581} (2000) 575--603.
%%CITATION = HEP-TH 0003098;%%

\bibitem{BBMS}
M.~Bertola, J.~Bros, U.~Moschella, R.~Schaeffer,
{\it A general construction of conformal field theories from scalar
anti-de Sitter quantum field theories}, 
Nucl.\ Phys.\ B {\bf 587} (2000) 619--644.
%%CITATION = NUPHA,B587,619;%%
%%CITATION = HEP-TH 9908140;%%

\bibitem{BF}
P.~Breitenlohner, D.~Z.~Freedman,
{\it Stability in gauged extended supergravity}, 
Ann.\ Physics {\bf 144} (1982) 249--281.
%%CITATION = APNYA,144,249;%% 

\bibitem{BEM} J. Bros, H. Epstein, U. Moschella, 
{\em Towards a general theory of quantized fields on the anti-de Sitter
space-time}, 
Commun.\ Math.\ Phys.\  {\bf 231} (2002) 481--528.
%%CITATION = HEP-TH 0111255;%%

\bibitem{B} H.-J. Borchers, 
{\em Field operators as $C^\infty$ functions in spacelike directions}, 
Nuovo Cim.\ {\bf 33} (1964) 1600--1613. 

\bibitem{D} V.~K.~Dobrev,
{\it Intertwining operator realization of the AdS/CFT correspondence},
Nucl.\ Phys.\ B {\bf 553} (1999) 559--582.
%%CITATION = HEP-TH 9812194;%%

\bibitem{DR} M.~D\"utsch, K.-H.~Rehren,
{\it A comment on the dual field in the AdS-CFT correspondence},
Lett.\ Math.\ Phys.\ {\bf 62} (2002) 171--184.
%%CITATION = HEP-TH 0204123;%%

\bibitem{GFF} M. D\"utsch, K.-H. Rehren: 
{\em Generalized free fields and the AdS-CFT correspondence},
Ann.\ Henri Poinc.\ {\bf 4} (2003) 613--635.
%%CITATION = MATH-PH 0209035;%%

\bibitem{PTGFF} M. D\"utsch, K.-H. Rehren: in preparation. 

\bibitem{FMMR} 
D.~Z.~Freedman, S.~D.~Mathur, A.~Matusis, L.~Rastelli,
{\it Correlation functions in the CFT$_d$/AdS$_{d+1}$ correspondence},
Nucl.\ Phys.\ B {\bf 546} (1999) 96--118; \\
%%CITATION = HEP-TH 9804058;%%
D.~Z.~Freedman, S.~D.~Mathur, A.~Matusis, L.~Rastelli,
{\it Comments on 4-point functions in the CFT/AdS correspondence},
Phys.\ Lett.\ B {\bf 452} (1999) 61--68.
%%CITATION = HEP-TH 9808006;%%

\bibitem{F} C.~Fronsdal,
{\it Elementary particles in a curved space. II},
Phys.\ Rev.\ D {\bf 10} (1974) 589--598.
%%CITATION = PHRVA,D10,589;%%

\bibitem{GJ} J. Glimm, A. Jaffe: {\em Quantum Physics: A Functional
    Point of View}, Springer Verlag, 1981.

\bibitem{AG} The rank $\leq 2$ cases are treated in: A. Grundmeier, 
    {\em Die Funktionalintegrale der AdS-CFT-Korrespondenz}, 
Diploma thesis, Univ.\ G\"ottingen (2004). 

\bibitem{GKP}
S.~S.~Gubser, I.~R.~Klebanov, A.~M.~Polyakov,
{\it Gauge theory correlators from non-critical string theory},
Phys.\ Lett.\ B {\bf 428} (1998) 105--114.
%%CITATION = HEP-TH 9802109;%%

\bibitem{J} A. Jaffe: {\em Constructive quantum field theory}
in: Mathematical Physics 2000, A. Fokas et al.\ (eds.), Imperial
College Press, London, 2000.

\bibitem{Jo}
R.~Jost, 
{\em The General Theory of Quantized Fields},
AMS, Providence, RI, 1965.

\bibitem{OK} O. Kniemeyer,  
{\em Untersuchungen am erzeugenden Funktional der AdS-CFT-Korrespondenz}, 
Diploma thesis, Univ.\ G\"ottingen (2002)

\bibitem{M} J.~M.~Maldacena,
{\it The large $N$ limit of superconformal field theories and supergravity},
Adv.\ Theor.\ Math.\ Phys.\  {\bf 2} (1998) 231--252.
%%CITATION = HEP-TH 9711200;%%

\bibitem{OS} K. Osterwalder, R. Schrader, 
{\em Axioms for Euclidean Green's functions, I+II},
Commun.\ Math.\ Phys.\  {\bf 31} (1973) 83--112, 
Commun.\ Math.\ Phys.\  {\bf 42} (1975) 281--305.
%%CITATION = CMPHA,31,83;%%
%%CITATION = CMPHA,42,281;%%

\bibitem{P} K. Pohlmeyer, 
{\em Eine scheinbare Abschw\"achung der Lokalit\"atsbedingung}, Commun.\
Math.\ Phys.\ {\bf 7} (1968) 80--92.

\bibitem{AH} 
K.-H.~Rehren,
{\it Algebraic holography}, 
Ann.\ Henri Poinc.\ {\bf 1} (2000) 607--623; \\
%%CITATION = HEP-TH 9905179;%%
K.-H.~Rehren,
{\it Local quantum observables in the AdS-CFT correspondence}, 
Phys.\ Lett.\ B {\bf 493} (2000) 383--388. 
%%CITATION = HEP-TH 0003120;%%

\bibitem{S}
R. Streater, {\em Outline of axiomatic relativistic quantum field
  theory}, Repts.\ Prog.\ Phys.\ {\bf 38} (1975) 771--846.

\bibitem{SW} 
R. Streater, A.S. Wightman, 
{\em PCT, Spin-Statistics, and All That}, 
Benjamin, New York, 1964. 

\bibitem{W}
E.~Witten,
{\it Anti-de Sitter space and holography},
Adv.\ Theor.\ Math.\ Phys.\ {\bf 2} (1998) 253--291.
%%CITATION = HEP-TH 9802150;%%

\end{footnotesize} 
\end{thebibliography}
\end{document}